%% file: main.tex
\documentclass{vgtc}                          % final (conference style)
\ifpdf%                                % if we use pdflatex
  \pdfoutput=1\relax                   % create PDFs from pdfLaTeX
  \usepackage{graphicx}                % allow us to embed graphics files
  \DeclareGraphicsExtensions{.pdf,.png,.jpg,.jpeg} % for pdflatex we expect .pdf, .png, or .jpg files
\else%                                 % else we use pure latex
  \usepackage{graphicx}                % allow us to embed graphics files
  \DeclareGraphicsExtensions{.eps}     % for pure latex we expect eps files
\fi%

\graphicspath{{figures/}{pictures/}{images/}{./}} % where to search for the images

\usepackage{microtype}                 % use micro-typography (slightly more compact, better to read)
\PassOptionsToPackage{warn}{textcomp}  % to address font issues with \textrightarrow
\usepackage{textcomp}                  % use better special symbols
\usepackage{mathptmx}                  % use matching math font
\usepackage{times}                     % we use Times as the main font
\usepackage{cite}                      % needed to automatically sort the references
\usepackage{tabu}                      % only used for the table example
\usepackage{booktabs}                  % only used for the table example
\usepackage{enumitem}
\usepackage[dvipsnames]{xcolor}

\onlineid{0}

\vgtccategory{Research}

\vgtcinsertpkg

\title{NeuroMapper: In-browser Visualizer for Neural Network Training}

\newcommand{\authorgap}{\hspace{10pt}}
\author{
  Zhiyan Zhou$^\ast$ %
   \authorgap
  Kevin Li$^\ast$
  \authorgap
  Haekyu Park$^\ast$
  \authorgap
  Megan Dass$^\ast$
  \authorgap
  Austin Wright$^\ast$
  \authorgap
  Nilaksh Das$^\ast$
  \authorgap
  Duen Horng Chau
  \thanks{\textsuperscript{\textrm 1}Georgia Institute of Technology. \{\href{mailto:zzhou406@gatech.edu}{zzhou406}$\mid$\href{mailto:kevin.li@gatech.edu}{kevin.li}$\mid$\href{mailto:haekyu@gatech.edu}{haekyu}$\mid$
  \href{mailto:mdass3@gatech.edu}{mdass3}$\mid$\href{mailto:apwright@gatech.edu}{apwright}$\mid$\href{mailto:nilakshdas@gatech.edu}{nilakshdas}$\mid$\href{mailto:polo@gatech.edu}{polo}\}@gatech.edu}
  \authorgap
}

\definecolor{applegreen}{rgb}{0.55, 0.71, 0.0}

\teaser{
  \centering
  \includegraphics[width=0.95\linewidth]{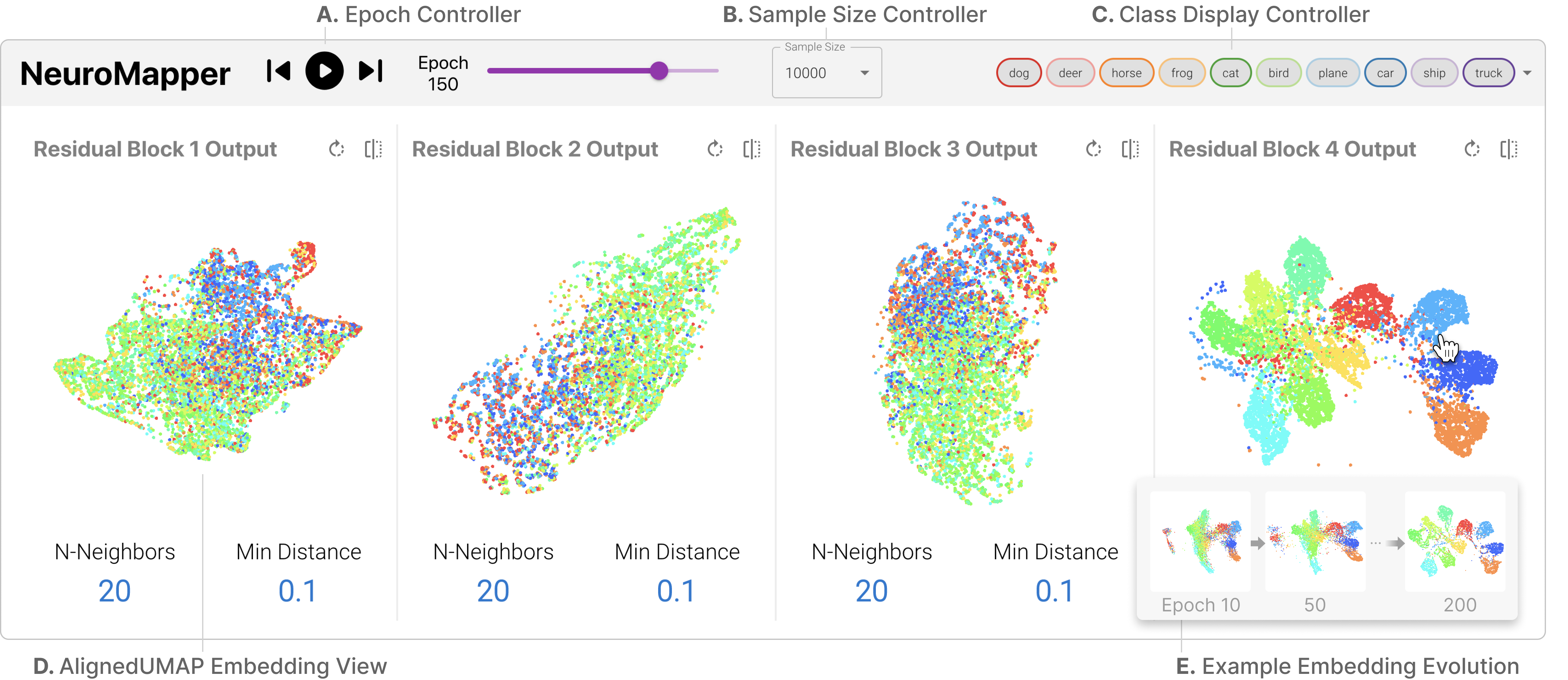}
  \caption{
    \method{} visualizes how a deep neural network (DNN) model, Resnet-50 trained on CIFAR10 dataset in this example, evolves during training by displaying the embeddings for training samples at the model's blocks  throughout the training regiment.
    \textbf{A. Epoch Controller} for updating the embedding display by adjusting the training epoch through a slider or auto-incrementation using the controller.
    \textbf{B. Sample Size Controller} changes the number of samples.
    \textbf{C. Class Display Controller} toggles classes' sample visibility.
    \textbf{D. AlignedUMAP Embedding View} visualizes input embeddings represented by each block in the DNN model, with user-adjustable hyperparameters at the bottom.
    \textbf{E. Example Embedding Evolution} of the fourth residual block of the model at epochs 10, 50, and 200.
  }
  \label{fig:teaser}
}

\abstract{
    \input{contents/Abstract}
} % end of abstract

\CCScatlist{
  \CCScatTwelve{Human-centered computing}{Visu\-al\-iza\-tion}{Visu\-al\-iza\-tion systems and tools}{};
}

\usepackage{xspace}
\newcommand{\method}[0]{\textsc{NeuroMapper}\xspace}
\definecolor{linkColor}{RGB}{6,125,233}
\begin{document}


\firstsection{Introduction}

\maketitle
\input{contents/Introduction}

\section{Designing \method}

\input{contents/SystemImplementation}

\section{Conclusion and Ongoing Work}

\input{contents/ConclusionAndFutureWork}

% The authors wish to thank A, B, and C. This work was supported in part by
% a grant from XYZ.}

%\bibliographystyle{abbrv}
\bibliographystyle{abbrv-doi}

\bibliography{template}
\end{document}

%% file: contents/Abstract.tex
We present our ongoing work \method, an in-browser visualization tool that helps machine learning (ML) developers interpret the evolution of a model during training, providing a new way to monitor the training process and visually discover reasons for suboptimal training.
While most existing deep neural networks (DNNs) interpretation tools are designed for already-trained model, 
\method scalably visualizes the evolution of the embeddings of a model's blocks 
across training epochs, enabling real-time visualization of 40,000 embedded points.To promote the embedding visualizations' spatial coherence across epochs, 
\method adapts \textit{AlignedUMAP}, a recent nonlinear dimensionality reduction technique to align the embeddings. With \method, users can explore the training dynamics of a Resnet-50 model, and adjust the embedding visualizations' parameters in real time. \method is open-sourced at \textcolor{linkColor}{\url{https://github.com/poloclub/NeuroMapper}} and runs in all modern web browsers. A demo of the tool in action is available at: \textcolor{linkColor}{\url{https://poloclub.github.io/NeuroMapper/}}.

%% file: contents/Introduction.tex
Deep neural networks (DNNs) are now widely used in many domains, increasing
the needs for understanding how they make decisions \cite{wagner2019interpretable}.
While some existing interpretation approaches use embeddings to help people interpret DNNs, they are often designed for 
providing insight within a single epoch \cite{zhong2017evolutionary}.

To scalably encode relationships of features and their dynamic changes during training while keeping the representation of samples across adjacent training epochs similar, we adapt a recent dimensionality reduction algorithm called AlignedUMAP 
\cite{alignedumapdoc}. AlignedUMAP not only distills the features of data samples into a low dimensional representation but also aligns the input embedding across training epochs so that the embedding of the same input data at adjacent epochs are more likely to be projected to similar vicinity,
such that model embedding changes induced by training may be more easily discovered through the visualization. 
Naively applying UMAP\cite{mcinnes2018umap-software} (or other nonlinear methods like t-SNE) for each epoch independently might lead to significant misalignment of embeddings, 
resulting in \textit{data-visual discrepancies} where underlying data changes may not be easily inferred from changes in the visualization \cite{li2020visualizing}.

In our work, we contribute the following:
\begin{itemize}[topsep=1pt, itemsep=0mm, parsep=3pt, leftmargin=9pt]
    \item 
    \textbf{\method{}, a cross-platform, browser-based tool that helps machine learning (ML) developers 
    visualize the training process of deep neural networks (DNNs).} 
    \method{} provides a new way to monitor and interpret the model evolution during training, going beyond only visualizing performance metrics (e.g., accuracy, F1-score) or interpreting already-trained models 
    ~\cite{smilkov2016embedding}.
    \item 
    \textbf{Open-sourced visualization that scales to 40,000 embedding points in real time.}
    \method takes the advantage of WebGL technology to drastically improve the visualization's rendering speed for large-scale embedding data. \method's open-sourced code-base and demo is available at \textcolor{linkColor}{\url{https://github.com/poloclub/NeuroMapper}}.  
    \item
    \textbf{Aligned visualizations for high-dimensional model embedding evolution.}
    Empowered by AlignedUMAP, \method visualizes the evolution of data representation by aligning the embedding of the same data points across training epochs. 
    AlignedUMAP is a nonlinear approach and has potential to better represent high-dimensional data points \cite{Maaten2008VisualizingDU}, providing a complementary technique to recent linear approaches \cite{li2020visualizing}.
    Users can adjust AlignedUMAP's hyperparameters in real time.
\end{itemize}

%% file: contents/SystemImplementation.tex
\smallskip
\noindent
\textbf{Aligning Input Data Representation Across Training Epochs.}
Visualizing embeddings over training epochs presents unique challenges, because the naive approach of independently generating and visualizing each epoch's embedding could lead to vastly different-looking visualizations at different epochs, where one epoch's embedding could be non-linearly scaled, rotated (or even flipped) very differently than another epoch's \cite{li2020visualizing}. 
Conceptually, since the embedding of a later epoch evolves from an earlier epoch's, 
the two epochs' embeddings should share some spatial resemblance. 
To accomplish such spatial coherence, we adapt AlignedUMAP \cite{alignedumapdoc} to align the embeddings of input in adjacent training epochs.
For each layer and epoch, \method first computes the high-demensional embedding of each input sample using the activation vector in the layer of the model at the training epoch.
\method then uses AlignedUMAP to align the 2D projection of such high-dimensional embeddings so that the embedding of the same input data at adjacent training epochs are projected on a similar location.
Specifically, AlignedUMAP optimizes a dimensionality reducer for each epoch at the same time while adding a constraint on the distance between the embeddings of the same data at adjacent epochs \cite{mcinnes2018umap-software}.
\smallskip
\noindent
\textbf{Open-sourced In-browser Scalable Visualization.}
Developed using modern web technologies such as React, MobX, and ScatterGL, \method's interface allows users to interpret how 10,000 samples from the CIFAR-10 dataset are encoded in a Resnet-50 model and how the embeddings evolve during the training process. Users can access the tool with any modern web-browser.

\begin{figure}[t]
    \centerline{
    \includegraphics[width=0.95\linewidth]{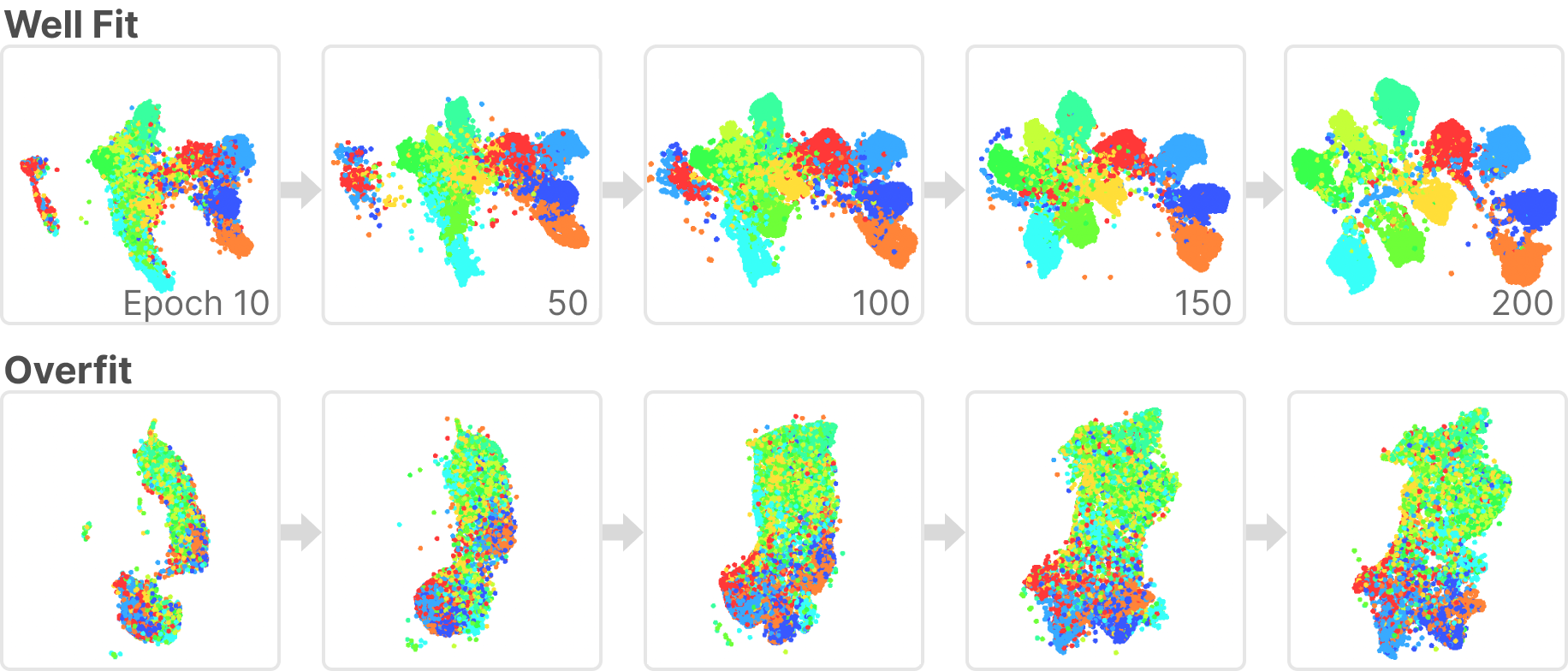}}
    \setlength{\belowcaptionskip}{-10pt}
    \caption{Evolution of embeddings at the fourth block of a well-fit and over-fit Resnet-50 model throughout training epochs.}
    \label{fig:fit}
\end{figure}

\smallskip
\noindent
\textbf{User Interface Design and Usage Scenario.}
The main view of \method shows how data representation evolves during training by visualizing the aligned 2D projections of the input data across training epochs.
By visualizing the evolution of data representation, users can interpret how a model is trained (e.g., whether it is trained well or not).
For instance, as seen in \autoref{fig:fit}, \method may help inform users whether a model is best-fit, under-fit, or over-fit by using the \textbf{Embedding View} (\autoref{fig:teaser}D) to visualize the aligned embedding of all ten thousands of data samples at the fourth block of a Resnet-50 model.
Classes of the input data
are differentiated by their color, and their display can be toggled in \textbf{Class Display} (\autoref{fig:teaser}C). 
To inspect the embedding, users can zoom and pan in the view.
When users hover over a data point in a layer, the same data point in all different layers is highlighted, and its class is displayed next to the highlighted points.
With the \textbf{Epoch Controller} (\autoref{fig:teaser}A) that looks like a typical video player, users can control which epoch to visualize.
Users can either use the auto-incrementation functionality by clicking on the start/pause button or manually adjust the epoch by dragging the slider or pressing the arrow buttons.
As users walk through the training epochs, they could notice the difference in separation between best-fit and over-fit models. For the best-fit model, there are clear separations between input embeddings of different classes, which is expected as block 4 is close to the output layer.
For the over-fitted model, there are multiple further separations within a cluster of the data of the same class,
meaning the model may overly differentiate the data of the same class.
To further inspect different aspects of model evolution, 
users can adjust the \textbf{AlignedUMAP Hyperparameters} (\autoref{fig:teaser}D)
such as \textit{N-neighbors} and \textit{Min Distance}.
\textit{N-neighbors} are the number of neighbors that AlignedUMAP look at when learning the input embeddings \cite{McInnes2018UMAPUM}.
For example, if users set lower values for \textit{N-neighbors}, they can focus more on the changes in the fine details of the feature similarity between the input data instead of how the bigger picture of the data representation evolves; lower values of \textit{N-neighbors} mean that AlignedUMAP focuses more on local neighborhoods of input data.
\textit{Min Distance} controls how tightly the embedding points are packed together \cite{McInnes2018UMAPUM}.
Users can adjust the number of samples displayed in each panel through the \textbf{Sample Size Controller} (\autoref{fig:teaser}B)

%% file: contents/ConclusionAndFutureWork.tex
\method is an open-sourced interactive visualization tool for interpreting how intermediate representation of input evolve during training. Using WebGL, \method can visualize large datasets in the modern browsers across multiple platforms. \method leverages AlignedUMAP to show a continuous evolution of intermediate output during training. We are working on extending \method to generate the input embedding in real time while a model is trained, so \method runs alongside the training process. Our goal is to apply \method to a wider range of models and datasets; it can handle visualizing tens of thousands of embedding points in real time, and we are planning to improve its scalability even more to handle more complex models and larger datasets through sampling. We also plan to improve \method's feature alignment by exploring Parametric UMAP, which utilizes a neural network to learn the relationship between data points and generate their embeddings. 

%% file: main.bbl
\begin{thebibliography}{1}

\bibitem{alignedumapdoc}
Aligned umap basic usage.
\newblock
  \url{https://umap-learn.readthedocs.io/en/latest/aligned_umap_basic_usage.html}.
\newblock Accessed: 2022-06-26.

\bibitem{li2020visualizing}
M.~Li, Z.~Zhao, and C.~Scheidegger.
\newblock Visualizing neural networks with the grand tour.
\newblock {\em Distill}, 2020.
\newblock https://distill.pub/2020/grand-tour. doi: {{%
10\hspace{.1pt}\discretionary{.}{%
}{.}\hspace{.4pt}23915\discretionary{/}{%
}{/}distill\hspace{.1pt}\discretionary{.}{%
}{.}\hspace{.4pt}00025}}


\bibitem{McInnes2018UMAPUM}
L.~McInnes and J.~Healy.
\newblock Umap: Uniform manifold approximation and projection for dimension
  reduction.
\newblock {\em ArXiv}, abs/1802.03426, 2018.

\bibitem{mcinnes2018umap-software}
L.~McInnes, J.~Healy, N.~Saul, and L.~Grossberger.
\newblock Umap: Uniform manifold approximation and projection.
\newblock {\em The Journal of Open Source Software}, 3(29):861, 2018.

\bibitem{smilkov2016embedding}
D.~Smilkov, N.~Thorat, C.~Nicholson, E.~Reif, F.~B. Vi{\'e}gas, and
  M.~Wattenberg.
\newblock Embedding projector: Interactive visualization and interpretation of
  embeddings.
\newblock {\em arXiv preprint arXiv:1611.05469}, 2016.

\bibitem{Maaten2008VisualizingDU}
L.~van~der Maaten and G.~E. Hinton.
\newblock Visualizing data using t-sne.
\newblock {\em Journal of Machine Learning Research}, 9:2579--2605, 2008.

\bibitem{wagner2019interpretable}
J.~Wagner, J.~M. Kohler, T.~Gindele, L.~Hetzel, J.~T. Wiedemer, and S.~Behnke.
\newblock Interpretable and fine-grained visual explanations for convolutional
  neural networks.
\newblock In {\em Proceedings of the IEEE/CVF Conference on Computer Vision and
  Pattern Recognition}, pp. 9097--9107, 2019.

\bibitem{zhong2017evolutionary}
W.~Zhong, C.~Xie, Y.~Zhong, Y.~Wang, W.~Xu, S.~Cheng, and K.~Mueller.
\newblock Evolutionary visual analysis of deep neural networks.
\newblock In {\em ICML Workshop on Visualization for Deep Learning}, p.~9,
  2017.

\end{thebibliography}
